\documentclass[a4paper,11pt]{article}
\usepackage{jheppub}

\title{\boldmath Localization of Matter Fields in the 5D Standing Wave Braneworld}

\author[a,b]{Merab Gogberashvili}
\affiliation[a]{Andronikashvili Institute of Physics,\\6 Tamarashvili Street, Tbilisi 0177, Georgia}
\affiliation[b]{Javakhishvili State University, Faculty of Exact and Natural Sciences,\\3 Chavchavadze Avenue, Tbilisi 0128, Georgia}
\emailAdd{gogber@gmail.com}

\abstract{We investigate the localization problem of matter fields within the 5D standing wave braneworld. In this model the brane emits anisotropic waves into the bulk with different amplitudes along different spatial dimensions. We show that in the case of increasing warp factor there exist the pure gravitational localization of all kinds of quantum and classical particles on the brane. For classical particles the anisotropy of the background metric is hidden, brane fields exhibit standard Lorentz symmetry in spite of anisotropic nature of the primordial 5D metric.}

\keywords{Field Theories in Higher Dimensions, Large Extra Dimensions, p-branes}

\arxivnumber{1204.2448 [hep-th]}

\begin{document}
\maketitle
\flushbottom

%%%%%%%%%%%%%%%%%%%%%%%%%%%%%%%%%%%%%%%%%%%%%%%%%%%%%%%%%%%%%%%%%%%%%%%%%%%%%%%
\section{Introduction}

Braneworld models involving large extra dimensions \cite{Hi-1,Hi-2,brane-1,brane-2} have been very useful in addressing several open questions in modern physics (for reviews see \cite{reviews-1,reviews-2,reviews-3,reviews-4}). Most of these models were realized as time independent field configurations. However, there have appeared also several non-stationary braneworlds \cite{S-1,S-2,S-3,S-4}. Here we consider the scenario with time-dependent metric coefficients recently proposed in \cite{Wave-1,Wave-2}.

A key requirement for realizing the braneworld idea is that the matter fields be localized on the brane. For several reasons one would like to have a universal gravitational trapping mechanism for all fields. However, there are difficulties to realize such a mechanism with exponentially warped space-times. In the existing 5D models the spin $0$ and spin $2$ fields can be localized on the brane with the decreasing warp factor \cite{brane-1,brane-2}, while the spin $1/2$ fields can be localized with the increasing factor \cite{BaGa}, and spin $1$ fields are not localized at all \cite{Po}. In the case of 6D models it was found that spin $0$, spin $1$ and spin $2$ fields are localized on the brane with the decreasing warp factor and spin $1/2$ fields again are localized with the increasing factor \cite{Od}. There exist also 6D models with non-exponential warp factors providing the gravitational localization of all kinds of bulk fields on the brane \cite{6D-1,6D-2,6D-3}, however, these models require an introduction of unnatural sources.

In this paper we show the universal gravitational trapping of all kinds of matter fields within the 5D standing waves braneworld \cite{Wave-1,Wave-2}. We start (Sect. 2) from recalling the basic ingredients of the model. In Sect. 3 and 4 we demonstrate the existence of zero modes of all kinds of quantum particles on the brane. The Sect. 5 is devoted to the geodesic motions of classical particles. A short summary and discussion are presented in Sect. 6.

%%%%%%%%%%%%%%%%%%%%%%%%%%%%%%%%%%%%%%%%%%%%%%%%%%%%%%%%%%%%%%%%%%%%

\section{Background solution}

In this section we briefly review the 5D standing wave braneworld model \cite{Wave-1,Wave-2}. The braneworld is generated by gravity coupled to a massless ghost scalar field, which depends on time and propagates in the bulk. In order to avoid the well-known problems of stability which occur with ghost fields, the bulk scalar field does not couple to ordinary matter in our model \cite{Wave-1,Wave-2}. For other models with phantom fields see \cite{phantom-1,phantom-2,phantom-3,phantom-4,phantom-5}.

The action of the model has the form:
\begin{equation} \label{action}
S = \int d^5x \sqrt{g} \left[\frac{1}{16 \pi G} \left( R - 2\Lambda\right) + g^{MN}\partial_M \phi \partial_N\phi \right]~,
\end{equation}
where the capital Latin indexes refer to 5D space-time and $G$ and $\Lambda$ are the 5D Newton and cosmological constants, respectively.

We use the metric {\it ansatz}:
\begin{equation} \label{metric}
ds^2 = e^{2a|r|}\left( dt^2 - e^{u}dx^2 - e^{u}dy^2 - e^{-2u}dz^2 \right) - dr^2~,
\end{equation}
where $a$ is a curvature scalar, with the determinant:
\begin{equation} \label{determinant}
\sqrt g = e^{4a|r|}~.
\end{equation}

The system of coupled Einstein and the scalar field equations for the {\it ansatz} (\ref{metric}) is self-consistent if:
\begin{equation} \label{tuning}
u(t,r) = \sqrt{\frac{8}{3} \pi G} ~ \phi (t,r) ~, ~~~~\Lambda = 6 a^2~,
\end{equation}
and it has the standing wave solution \cite{Wave-1,Wave-2}:
\begin{equation} \label{u}
u(t,r) = \sin (\omega t) Z(r)~,
\end{equation}
where
\begin{equation} \label{Z}
Z(r) = C e^{-2a|r|} J_2\left( \frac{\omega}{a} e^{-a|r|} \right)~.
\end{equation}
Here $C$ and $\omega$ are real constants and $J_2$ is the second-order Bessel function of the first kind. Thus the model (\ref{metric}) describes the brane located at $r = 0$ which possesses the anisotropic oscillations and sends a wave into the bulk (as in \cite{GMS-1,GMS-2}), i.e. the brane is warped along the spatial coordinates through the factors $\sim e^{u(t,r)}$, which depend on time $t$ and the extra coordinate $r$.

As pointed out in \cite{Wave-1,Wave-2}, the ghost-like field $\phi (t,r)$, along with the metric oscillations $u(t,r)$, must be unobservable on the brane. According to (\ref{u}) we can accomplish this requirement by setting the boundary condition for the Bessel function:
\begin{equation}
\left. J_2(r) \right|_{r=0} = 0~.
\end{equation}
Since $J_2$ is an oscillatory function, in the case of increasing (decreasing) warp factor the function (\ref{Z}) can have finite (infinite) number of zeros. Thus the above boundary condition can be written in the form which quantizes the oscillation frequency $\omega$ of the standing wave in terms of the curvature scale $a$, i.e.
\begin{equation} \label{quantize}
\frac{\omega}{|a|} = X_n~,
\end{equation}
where $X_n$ is the $n^{th}$ zero of $J_2(r)$. Correspondingly, the nodes of the standing wave in the bulk, the points where the functions $\phi (t,r)$ and $u(t,r)$ vanish, can be considered as 4D space-time 'islands', where the matter particles are assumed to be bound.

We consider the case with the increasing warp factor ($a>0$) when the function $J_2( e^{-a|r|})$ in (\ref{u}) has a finite number of zeros. For simplicity, in this paper we explore the case when $J_2$ has a single zero at the position of the brane, $r = 0$, i.e. we assume that
\begin{equation}\label{FirstZerosY}
\frac{\omega}{a} = X_1 \approx 5.14~.
\end{equation}

In the equations of matter fields the oscillatory function (\ref{u}) enters via some exponential functions:
\begin{equation} \label{e-br}
e^{bu} = \sum \limits_{n = 0}^{+\infty } \frac{\left( bu \right)^n}{n!}~,
\end{equation}
where $b$ is a constant. We suppose that the frequency $\omega$ of standing waves in the oscillatory metric function $u(t,r)$ is much larger than the frequencies associated with the energies of particles on the brane. In this case we can perform the time averaging of oscillating exponents in the equation of the matter fields.

Using the expression:
\begin{equation}
\frac{\omega}{2\pi} \int\limits_0^{2\pi/\omega} \sin^m (\omega t) dt = \left \{
\begin{array} {lr}
0 & ( m = 2n + 1)\\
\frac {m!}{2^m(m/2)!^2} & (m = 2n)
\end{array}
\right.
\end{equation}
we find that the time averages of some oscillatory functions are zero \cite{GMM-1}:
\begin{equation}\label{AdditionalFacts}
\left\langle u \right\rangle = \left\langle u' \right\rangle= \left\langle \dot{u} \right\rangle =\left\langle \left( e^{ bu}\right)^\cdot \right\rangle = 0~,
\end{equation}
where the primes and dots mean the derivatives with respect to the extra coordinate $r$ and time coordinate $t$, respectively. We also obtain the simple formulas for the non-vanishing averages:
\begin{eqnarray} \label{e-u}
\left\langle e^{bu} \right\rangle = \sum \limits_{n = 0}^{+\infty } \frac{ (bZ)^{2n}}{2^{2n}n!^2} = I_0 ~, \nonumber \\
\left\langle \left(e^{bu}\right)' \right\rangle = bZ'\sum \limits_{n = 0}^{+\infty } \frac{ (bZ)^{2n+1}}{2^{2n}(2n+1)n!^2} = b^2ZZ'\left[ I_0 + \frac{\pi}{2}\left(I_0L_1-I_1L_0\right) \right]~,
\end{eqnarray}
where $I_0$, $I_1$ and $L_0$, $L_1$ are the modified Bessel and Struve functions of the argument $bZ$, respectively.

%%%%%%%%%%%%%%%%%%%%%%%%%%%%%%%%%%%%%%%%%%%%%%%%%%%%%%%%%%%%%%%%%%%%%%%%

\section{Localization of bosons}

Let us consider the localization problem of quantum fields with spin $0$, $1$ and $2$ \cite{GMM-1, GMM-2}.

The massless scalar field can be defined by the 5D action:
\begin{equation} \label{Sphi}
S_\Phi = - \frac 12 \int d^5x\sqrt{g} ~ g^{MN}\partial_M\Phi \partial_N\Phi~.
\end{equation}
The corresponding Klein-Gordon equation,
\begin{equation}\label{ScalFieldEqn}
\frac{1}{\sqrt g}~\partial_M \left( \sqrt g g^{MN}\partial_N \Phi \right) = 0~,
\end{equation}
on the background (\ref{metric}) has the form:
\begin{eqnarray}\label{Equation1}
\left[ \partial_t^2 - e^{- u}( \partial_x^2 + \partial_y^2) - e^{2u} \partial_z^2\right]\Phi = e^{2a|r|}\left( e^{4a|r|} \Phi' \right)'.
\end{eqnarray}
We look for the solution of this equation in the form:
\begin{equation}\label{Solution1}
\Phi \left( {x^\nu,r} \right) = e^{ip_\nu x^\nu} \xi (r)
\end{equation}
(Greek letters are used for 4D indices), which transforms (\ref{Equation1}) into the equation,
\begin{equation} \label{xi-1}
\left( e^{4a|r|}\xi' \right)' = e^{2a|r|}\left[(p_x^2 + p_y^2)e^{-u} + p_z^2 e^{2u} - p_t^2\right]\xi~.
\end{equation}

When the frequency $\omega$ of standing waves is much larger than the frequencies associated with the energy $p_t$ of particles on the brane we can replace the oscillatory exponents in (\ref{xi-1}) by their time averages (\ref{e-u}). Then we obtain the equation for the extra factor $\xi (r)$:
\begin{equation}\label{xi-2}
\left( e^{4a|r|}\xi' \right)' - e^{2a|r|}P^2(r) \xi = 0~,
\end{equation}
which contains the {\it $r$-dependent momentum}:
\begin{equation} \label{P2}
P^2(r) = \left(\left\langle e^{-u} \right\rangle -1\right)\left( p_x^2 + p_y^2 \right) + \left(\left\langle e^{2u} \right\rangle -1\right)p_z^2 ~.
\end{equation}

On the brane, where $u\approx 0$, the parameters $p_\nu$ can be regarded as components of energy-momentum along the brane which obeys the dispersion relation:
\begin{equation}
p_t^2 - p_x^2 - p_y^2 - p_z^2 = 0~.
\end{equation}

To study a general behavior of the extra part of the scalar zero mode wave function we explore (\ref{xi-2}) in two limiting regions: far from and close to the brane, respectively. The {\it $r$-dependent momentum} (\ref{P2}), which describes oscillatory properties of standing waves, has the following asymptotical forms:
\begin{eqnarray}
\left. P^2 (r)\right|_{r \to 0} &\sim& r^2~, \nonumber \\
\left. P^2 (r)\right|_{r \to \infty} &\sim& const~.
\end{eqnarray}
Then we solve (\ref{xi-2}) in this limiting regions and find:
\begin{eqnarray} \label{solution-s}
\left. \xi (r)\right|_{r \to 0} &\sim& const ~, \nonumber \\
\left. \xi (r)\right|_{r \to \infty} &\sim& e^{-4a|r|}~.
\end{eqnarray}
Hence $\xi (r)$ has maximum on the brane and falls off at the infinity as $e^{-4a|r|}$.

In the action of scalar fields (\ref{Sphi}) the determinant (\ref{determinant}) and the metric tensor with upper indices give the total exponential factor $e^{2a|r|}$, which obviously increases for $a > 0$. This is the reason why in the original brane models \cite{brane-1,brane-2} the scalar field zero modes with the constant $r$-dependent extra part can be localized on the brane only in the case of decreasing warp factor (i.e. $a<0$). In our model the extra part of wave function (\ref{solution-s}) is not constant, moreover, for $a > 0$ it contains the exponentially decreasing factor $e^{-4a|r|}$. For such an extra dimension factor the integral over $r$ in the action (\ref{Sphi}) is convergent, hence 4D scalar fields are localized on the brane \cite{GMM-1}.

It is known that the transverse traceless graviton modes obey the equation of a massless scalar field in a curved background. Indeed, let us consider the metric fluctuations:
\begin{equation} \label{metric-h}
ds^2 = e^{2a|r|}\left( g_{\mu\nu} + h_{\mu\nu}\right)dx^\mu dx^\nu - dr^2~,
\end{equation}
where $g_{\mu\nu}$ is the metric tensor of the 4D part of (\ref{metric}):
\begin{equation} \label{metric-4}
g_{\mu\nu} = \left( 1, - e^{u}, - e^{u}, - e^{-2u} \right) ~.
\end{equation}

Close to the brane ($u\approx 0$) for time averages of oscillatory exponents (\ref{e-u}) we can use the approximation:
\begin{equation}
\left\langle e^{u(t,r)} \right\rangle \approx 1 + \left\langle u \right\rangle~.
\end{equation}
Thus the functions $\left\langle u \right\rangle$ can be regarded as $r$-dependent additive terms of $h_{\mu\nu}$. Then the equations of motion for the fluctuations $h_{\mu\nu}$:
\begin{equation}\label{graviton}
\frac{1}{\sqrt g}~\partial_M \left( \sqrt g g^{MN}\partial_N h_{\mu\nu} \right) = 0~,
\end{equation}
are equivalent to the equation of motion of a scalar field (\ref{ScalFieldEqn}) if we replace $\Phi$ with $h_{\mu\nu}$. Accordingly, the condition of localization of spin-$2$ graviton field might be equivalent to that of spin-$0$ scalar field considered above.

A little bit more complicated task is to show the localization of vector field zero modes \cite{GMM-2}. Let us consider only the $U(1)$ vector field (the generalization to the case of non-Abelian gauge fields is straightforward).

The 5D action of vector fields,
\begin{equation}\label{VectorAction}
S_A = - \frac14\int d^5x\sqrt g~ g^{MN}g^{PR}F_{MP}F_{NR}~,
\end{equation}
where
\begin{equation} \label{F}
F_{MP} = \partial _M A_P - \partial _P A_M~,
\end{equation}
leads to the system of five equations:
\begin{equation}\label{VectorFieldEquations}
\frac{1}{\sqrt g}\partial _M\left( \sqrt g ~g^{MN}g^{PR}F_{NR}
\right) = 0 ~.
\end{equation}

We seek for the solution of the system (\ref{VectorFieldEquations}) in the form:
\begin{eqnarray}\label{VectorDecomposition}
A_t(x^C) &=& \upsilon (r)~a_t(x^\nu)~, \nonumber \\
A_x(x^C) &=& e^{u(t,r)} \upsilon (r)~a_x(x^\nu)~, \nonumber \\
A_y(x^C) &=& e^{u(t,r)} \upsilon (r)~a_y(x^\nu)~, \\
A_z(x^C) &=& e^{-2u(t,r)} \upsilon (r)~a_z(x^\nu)~, \nonumber \\
A_r(x^C) &=& 0~,\nonumber
\end{eqnarray}
where $a_{\mu}(x^\nu)$ denote the components of 4D vector potential and scalar factor $\upsilon (r)$ depends only on the extra coordinate $r$.

Taking into account the equalities (\ref{AdditionalFacts}), the time averaging of the fifth equation of the system (\ref{VectorFieldEquations}),
\begin{equation}
\partial_\alpha \left( g^{\alpha\beta} A_\beta '\right) = 0~,
\end{equation}
yields the Lorenz-like gauge condition:
\begin{equation} \label{GaugeCondition2}
g^{\alpha\beta}\partial_\alpha A_\beta = \eta^{\alpha\beta}\partial_\alpha a_\beta = 0~,
\end{equation}
where $\eta_{\alpha\beta}$ denotes the metric of 4D Minkowski space-time. The equation (\ref{GaugeCondition2}), together with the last expression of (\ref{VectorDecomposition}), can be considered as the full set of imposed gauge conditions.

The remaining four equations of the system (\ref{VectorFieldEquations}),
\begin{equation}
\partial _\gamma\left( g^{\gamma\delta}g^{\beta\alpha}F_{\delta\alpha} \right) - \frac{1}{\sqrt g }\left( \sqrt g ~g^{\beta\alpha}A'_\alpha \right)' = 0 ~,
\end{equation}
after the time averaging and the use of (\ref{AdditionalFacts}) and (\ref{GaugeCondition2}) reduce to:
\begin{equation} \label{system-a}
\upsilon ~g^{\alpha \delta}\partial _\alpha\partial _\delta a_\beta + e^{-2a|r|} \left( e^{2a|r|}\upsilon' \right)' a_\beta = 0 ~.
\end{equation}
We require the existence of 4D vector waves localized on the brane,
\begin{equation} \label{Factors}
a_\mu\left(x^\nu\right) \sim \varepsilon_\mu e^{ip_\nu x^\nu} ~,
\end{equation}
where $p_\nu$ are the components of energy-momentum along the brane. Then the system (\ref{system-a}) yields the single equation for $\upsilon (r)$:
\begin{equation}\label{ups}
\left( e^{2a|r|}\upsilon ' \right)' - P^2(r) \upsilon  = 0~,
\end{equation}
where $P^2(r)$ is done by (\ref{P2}). The solutions of (\ref{ups}) close to and far from the brane are:
\begin{eqnarray} \label{v}
\left. \upsilon (r)\right|_{r \to 0} &\sim& const ~, \nonumber \\
\left. \upsilon(r)\right|_{r \to \infty} &\sim&  e^{-2 a|r|}~.
\end{eqnarray}
We see that the extra factor $\upsilon (r)$ of the vector field zero mode wave function has a maximum on the brane and falls off at the infinity as $e^{-2a|r|}$.

In 5D brane models the implementation of pure gravitational trapping mechanism of vector field particles remains the most problematic. The reason is that in the vector action (\ref{VectorAction}) the extra dimension parts of the determinant (\ref{determinant}) and two metric tensors with upper indices cancel each other. Because of this in the original brane models the vector field zero modes (with the constant extra dimension part) cannot be localized on the brane for any sign of $a$. In our model the extra dimension part of vector field $\upsilon (r)$ is not constant; at the infinity it decreases exponentially (\ref{v}). Thus the integral over $r$ in the action (\ref{VectorAction}) is convergent, what means that 4D vector fields are localized on the brane \cite{GMM-2}.

%%%%%%%%%%%%%%%%%%%%%%%%%%%%%%%%%%%%%%%%%%%%%%%%%%%%%%%%%%%%%%%%%%%%%%%%

\section{Localization of fermions}

Now we investigate the localization problem for massless fermions \cite{GMM-3}.

For Minkowskian $4\times 4$ gamma matrices ($\{ \gamma^\alpha, \gamma^\beta \} = 2\eta^{\alpha\beta}$) we use the Weyl basis:
\begin{equation}\label{MinkowskianGammaMatrices}
\begin{array}{l}
\gamma ^t =~ \left( {\begin{array}{*{20}{c}}
0&I\\
I&0
\end{array}} \right),~~~
{\gamma ^i} = \left( {\begin{array}{*{20}{c}}
0&-\sigma^i\\
\sigma^i&0
\end{array}} \right),
\end{array}
\end{equation}
where $I$ and $\sigma^i$ ($i = x,y,z$) denote the standard $2\times2$ unit and Pauli matrices respectively.

5D gamma matrices $\Upsilon^A = h_{\bar A}^A\Upsilon^{\bar A}$ can be chosen as:
\begin{eqnarray}\label{GammaMatricesRelation}
\Upsilon ^t &=& e^{ - a|r|}~\gamma ^t ~, \nonumber \\
\Upsilon ^x &=& e^{ - a|r| - u/2}~ \gamma ^x~, \nonumber \\
\Upsilon ^y &=& e^{ - a|r| - u/2}~\gamma ^y~, \\
\Upsilon ^z &=&  e^{ - a|r| + u}~ \gamma ^z~, \nonumber \\
\Upsilon ^r &=& i\gamma ^5~. \nonumber
\end{eqnarray}

The 5D Dirac action for massless fermions:
\begin{equation}\label{SpinorAction}
S_\Psi = \int d^5x \sqrt g ~i\overline \Psi \left(x^A\right) \Upsilon ^MD_M \Psi \left(x^A\right)~,
\end{equation}
contains the covariant derivatives:
\begin{equation}
D_A = \partial_A + \frac 14 \Omega_A^{\bar B \bar C} \Upsilon_{\bar B} \Upsilon_{\bar C}~.
\end{equation}
The non-vanishing components of the spin-connection in the background (\ref{metric}) are:
\begin{eqnarray}\label{Spin-ConnectionComponents}
\Omega_t^{\bar t \bar r} &=& - \left( e^{a|r|} \right)'~, \nonumber \\
\Omega_x^{\bar x \bar r} &=& \Omega_y^{\bar y \bar r} = - \left( e^{a|r| + u/2 } \right)'~, \nonumber \\
\Omega_z^{\bar z \bar r} &=& - \left( e^{a|r| - u} \right)' ~,\\
\Omega_x^{\bar x \bar t} &=& \Omega_y^{\bar y \bar t} = \left( e^{u/2}\right)^\cdot~, \nonumber \\
\Omega_z^{\bar z \bar t} &=& \left(  e^{ - u}\right)^\cdot~. \nonumber
\end{eqnarray}

The corresponding to (\ref{SpinorAction}) 5D Dirac equation reads:
\begin{equation}\label{SpinorEquation}
i\Upsilon^AD_A\Psi = i\left( \Upsilon ^\mu D_\mu + \Upsilon ^rD_r \right)\Psi = 0~.
\end{equation}
For wave function of the bulk fermion field we use the chiral decomposition:
\begin{equation}\label{Psi}
\Psi \left(x^\nu,r\right) = \psi_L \left(x^\nu\right) \lambda (r) + \psi_R \left(x^\nu\right) \rho (r)~,
\end{equation}
where $\lambda(r)$ and $\rho(r)$ are the extra dimension factors of the left and right fermion wave functions respectively. We assume that 4D left and right Dirac spinors:
\begin{equation} \label{chiral}
\gamma^5 \psi_L = - \psi_L~, ~~~~~
\gamma^5 \psi_R = + \psi_R ~,
\end{equation}
correspond to the zero mode wave functions, i.e. they satisfy the free Dirac equations:
\begin{equation} \label{Dirac-free}
i\gamma^\mu \partial_\mu \psi_L = i\gamma^\mu \partial_\mu \psi_R = 0~.
\end{equation}

The solutions of (\ref{Dirac-free}) in our representation (\ref{MinkowskianGammaMatrices}) can be written in the form:
\begin{eqnarray} \label{psi-free}
\psi_R (x^\nu) = \left( \begin{array}{c}R\\0\end{array} \right) e^{-ip_\nu x^\nu}, \nonumber \\
\psi_L (x^\nu) = \left( \begin{array}{c}0\\L\end{array} \right) e^{-ip_\nu x^\nu},
\end{eqnarray}
where the constant 2-spinors $L$ and $R$ satisfy:
\begin{equation} \label{relations}
\left(p_t + \sigma^ip_i\right)L = \left(p_t - \sigma^ip_i\right)R = 0~.
\end{equation}

When the frequency $\omega$ of standing waves is much larger than the frequencies associated with the energies $p_t$ of the fermions on the brane we can time average the oscillatory functions in the Dirac equation (\ref{SpinorEquation}). Time averages of the Dirac operators are:
\begin{eqnarray}\label{TimeAveragesOfDiracOperators}
\left\langle i\Upsilon^tD_t \right\rangle &=& ie^{ -a|r|}~\gamma ^t\partial_t - \frac 12 a ~sgn(r)\gamma^5, \nonumber \\
\left\langle i\Upsilon^xD_x \right\rangle &=& ie^{ -a|r|}\left\langle e^{ -u/2} \right\rangle \gamma ^x\partial_x - \frac 12 a ~sgn(r)\gamma^5, \nonumber \\
\left\langle i\Upsilon^yD_y \right\rangle &=& ie^{ - a|r|}\left\langle e^{ -u/2} \right\rangle \gamma ^y\partial_y - \frac 12 a ~sgn(r)\gamma^5, \nonumber \\
\left\langle i\Upsilon^zD_z \right\rangle &=& ie^{- a|r|}\left\langle e^{ u} \right\rangle \gamma ^z\partial_z - \frac 12 a ~sgn(r)\gamma^5, \\
\left\langle i\Upsilon^rD_r \right\rangle &=& -\gamma ^5\partial_r, \nonumber
\end{eqnarray}
and the equation (\ref{SpinorEquation}) takes the form:
\begin{equation}\label{SpinorEquation1}
i\left[{\gamma^t}{\partial _t} + \left\langle e^{ u/2} \right\rangle \left(\gamma ^x\partial _x + \gamma ^y\partial _y\right)  + \left\langle e^{-u} \right\rangle \gamma ^z\partial _z\right]\Psi = e^{a|r|}\gamma^5 \left[2a ~sgn(r) + \partial_r \right]\Psi~.
\end{equation}
Using the solutions of free equations (\ref{psi-free}) and the relations (\ref{relations}) it can be rewritten as the system:
\begin{eqnarray}\label{L-R}
\left( \begin{array}{*{20}{c}}
-e^{a|r|}\left[2a ~sgn(r) + \partial_r \right]&\sigma^i{\cal P}_i(r)\\
- \sigma^i{\cal P}_i(r)&e^{a|r|}\left[2a ~sgn(r) + \partial_r \right]
\end{array} \right)\left( \begin{array}{*{10}{c}} \rho (r)R\\
\lambda (r)L\end{array} \right) = 0~.
\end{eqnarray}
Here we have introduced the functions ${\cal P}_i (r)$:
\begin{eqnarray} \label{P-i}
{\cal P}_x (r) &=& \left(\left\langle e^{-u/2} \right\rangle -1\right) p_x = \left[I_0\left(|C|Z/2\right)-1\right] p_x, \nonumber \\
{\cal P}_y (r) &=& \left(\left\langle e^{-u/2} \right\rangle - 1 \right)p_y = \left[I_0\left(|C|Z/2\right)-1\right] p_y, \nonumber \\
{\cal P}_z (r) &=& \left(\left\langle e^{u} \right\rangle - 1 \right)p_z = \left[I_0\left(|C|Z\right)-1\right] p_z,
\end{eqnarray}
where $Z(r)$ is defined in (\ref{Z}). These functions, as (\ref{P2}), can be considered as the components of {\it '$r$-dependent momentum'} of the spinor field:
\begin{equation}\label{P}
{\cal P}^2(r) = {\cal P}_x^2 + {\cal P}_y^2 + {\cal P}_z^2~.
\end{equation}

From the second equation of the system (\ref{L-R}) it is straightforward to find
\begin{equation} \label{rho=lambda}
\rho (r) R = e^{a|r|} \frac {\sigma^i{\cal P}_i(r)}{{\cal P}^2(r)}\left[2a ~sgn(r) + \partial_r \right]\lambda (r) L~.
\end{equation}
Inserting (\ref{rho=lambda}) into the first equation of (\ref{L-R}) and multiplying the result by $\sigma^i{\cal P}_i$, we receive the second order differential equation for the function $\lambda (r)$:
\begin{equation}\label{L-Equation}
\lambda'' + \left[ 5a~ sgn( r ) - \frac{\cal P'}{\cal P}\right] \lambda' + \left[4a \delta (r) + 6a^2 - 2a~sgn(r)\frac{\cal P'}{\cal P} - {\cal P}^2e^{ - 2a|r|} \right] \lambda = 0~.
\end{equation}
Now, as for the case of bosonic fields, let us investigate this equation in the domains far from and close to the brane, respectively.

Close to the brane the {\it '$r$-dependent momentum'} (\ref{P}) behaves as:
\begin{equation}
\left. {\cal P} (r)\right|_{r\to \pm 0} = A r^2 + O(r^3)~,
\end{equation}
where $A$ is a constant, and the equation (\ref{L-Equation}) takes the following asymptotic form:
\begin{equation}
\lambda'' + \left[ 5a~ sgn( r ) - \frac{2}{r}\right] \lambda' + \left[4a \delta (r) + 6a^2 - \frac {4a}{r}~sgn(r) \right] \lambda = 0~.
\end{equation}
The main solution of this equation in our approximation is:
\begin{equation}\label{L-0}
\lambda (r)|_{r \to \pm 0} = B e^{ - 2a|r|}~,
\end{equation}
where $B$ is a constant.

Note that, as a consequence of (\ref{rho=lambda}) and (\ref{L-0}), in our setup the right fermionic modes are absent on the brane:
\begin{equation}\label{rho-0}
\rho (r)|_{r \to \pm 0} = 0 ~.
\end{equation}

In the second limited region - far away from the brane,
\begin{equation}
{\cal P} (r)|_{r \to \pm \infty} \sim const~,
\end{equation}
and the equation (\ref{L-Equation}) takes the asymptotic form:
\begin{equation}\label{L-infinity_0}
\lambda'' + 5a~ sgn( r ) \lambda' + 6a^2 \lambda = 0~,
\end{equation}
with the solution:
\begin{equation}\label{L-infinity}
\lambda (r)|_{r \to \pm \infty } \sim e^{ - 3a|r|}~.
\end{equation}
Using (\ref{L-infinity}) from the relation (\ref{rho=lambda}) we find also the asymptotic behavior of the extra dimension factor of the right fermion wave function:
\begin{equation}\label{R-infinity}
\rho (r)|_{r \to \pm \infty } \sim e^{ -2 a|r|}~.
\end{equation}

So in our model the extra dimension part of the left spinor wave function (\ref{Psi}) has the maximum at the origin,
\begin{equation}
\lambda (r)|_{r=0} = B~,
\end{equation}
decreases from the brane, and turns into the asymptotic form (\ref{L-infinity}) at the infinity. When $r\to \infty$ the determinant (\ref{determinant}) in the action integral (\ref{SpinorAction}) increases as $e^{4a|r|}$. However, {\it f\"{u}nfbein} in (\ref{TimeAveragesOfDiracOperators}) contribute $e^{-a|r|}$ and the extra dimension factor of left fermions (\ref{L-infinity}) is proportional to $e^{-6a|r|}$. Thus the overall $r$-dependent part decreases as $e^{-3a|r|}$, i.e. the integral over $r$ in (\ref{SpinorAction}) is convergent and zero modes of left fermions are localized on the brane.

Due to (\ref{R-infinity}) the extra dimension part of right fermions cancels the factor from the determinant (\ref{determinant}) in the action (\ref{SpinorAction}). The $r$-component of the Dirac operator (\ref{TimeAveragesOfDiracOperators}) does not contain an extra decreasing factor from {\it f\"{u}nfbein} and the integral over $r$ of the term $i\overline \Psi ~\Upsilon^r\partial_r \Psi$ in (\ref{SpinorAction}) diverges, i.e. the zero mode wavefunctions of right fermions actually are not normalizable.

%%%%%%%%%%%%%%%%%%%%%%%%%%%%%%%%%%%%%%%%%%%%%%%%%%%%%%%%%%%%%%%%

\section{Classical particles}

Finally we consider the motion of a classical particle, or a photon, which obey 5D geodesic equation of motion:
\begin{equation} \label{Geo}
\frac{d^2 x^A}{dk^2} + \Gamma^A_{BC} \frac{d x^B}{dk}\frac{d x^C}{dk} = 0~,
\end{equation}
where $k$ is the parameter of trajectory. Non-zero components of 5D Cristoffel symbols for the metric (\ref{metric}) are:
\begin{eqnarray} \label{Christoffel}
\Gamma^t_{tr} = a|r|', ~~\Gamma^t_{zz}= \frac 12 \left(e^{-2u}\right)^\cdot, ~~\Gamma^t_{xx}= \Gamma^t_{yy} = \frac 12 \left(e^u\right)^\cdot, \nonumber \\
\Gamma^x_{tx}= \Gamma^y_{ty}= \frac 12 u^\cdot, ~~\Gamma^z_{tz}= - u^\cdot, ~~\Gamma^x_{rx}= \Gamma^y_{ry}= a|r|'+\frac 12 u', ~~\Gamma^z_{rz}= a|r|'- u',\\
\Gamma^r_{tt} = \frac 12 \left(e^{2a|r|}\right)', ~~\Gamma^r_{zz} = -\frac 12 \left(e^{2a|r|-2 u}\right)', ~~\Gamma^r_{xx} = \Gamma^r_{yy} = -\frac 12 \left(e^{2a|r|+u}\right)'.\nonumber
\end{eqnarray}

For simplicity we shall consider the motion in $(rx)$-plane, i.e. we suppose
\begin{equation} \label{yz}
d y = d z = 0~.
\end{equation}
Then $y$ and $z$ are not dynamical variables and the system (\ref{Geo}) consists of three independent equations:
\begin{eqnarray} \label{Geo-sys}
\frac{d^2 t}{dk^2} + \Gamma^t_{tr} \frac{d t}{d k}\frac{d r}{d k} + \Gamma^t_{xx} \left(\frac{d x}{d k}\right)^2 = 0~, \nonumber \\
\frac{d^2 x}{dk^2} + \Gamma^x_{tx} \frac{d t}{d k}\frac{d x}{d k} + \Gamma^x_{rx} \frac{d r}{d k}\frac{d x}{d k} = 0~, \\
\frac{d^2 r}{dk^2} + \Gamma^r_{tt} \left(\frac{d t}{d k}\right)^2 + \Gamma^r_{xx} \left(\frac{d x}{d k}\right)^2 = 0~. \nonumber
\end{eqnarray}

The second equation of this system, after dividing by $dx/dk$, can be rewritten as:
\begin{equation}
\frac{d}{d k}\left[ \ln \left(\frac{d x}{d k}\right) + \frac {\dot{u}}{2} \frac{d t}{d k} + \left( a|r|' + \frac {u'}{2} \right) \frac{d r}{d k}\right] = 0~,
\end{equation}
It's first integral is:
\begin{equation} \label{dx}
\frac{dx}{dk} = V e^{-a|r|-u/2}~,
\end{equation}
where the constant $V$ corresponds to the component of the particles velocity along the brane.

Inserting (\ref{dx}) into the first equation of the system (\ref{Geo-sys}) we find:
\begin{equation}
\frac{d}{d k}\left[ \ln \left(\frac{d t}{d k}\right) + a|r|'\right] + \frac 12 \dot{u}V^2 e^{-2a|r|}\frac{d k}{d t}= 0~.
\end{equation}
In the case of fast oscillations of standing waves the time average of the last term yields zero (since $\left\langle \dot{u} \right\rangle = 0 $) and
\begin{equation} \label{dt}
\frac {d t}{d k}= e^{-a|r|} ~.
\end{equation}
The integration constant in this expression was included in the definition of $k$, so that on the brane ($r=0$) the parameter $k$ coincides with the coordinate time $t$.

Inserting (\ref{Christoffel}), (\ref{dx}) and (\ref{dt}) into the last equation of (\ref{Geo-sys}) and multiplying it by $dr/dk$ we find:
\begin{equation}
\frac{d}{d k}\left[ \frac 12 \left(\frac{d r}{d k}\right)^2 + a(1-V^2) |r| - \frac 12 V^2u \right] = 0~.
\end{equation}
Time average of the last term in this equation gives zero ($\left\langle u \right\rangle = 0 $) and the first integral is:
\begin{equation} \label{dr}
\frac 12 \left(\frac{d r}{d k}\right)^2 + a(1-V^2) |r| = \epsilon~,
\end{equation}
where the constant $\epsilon >0$ corresponds to the energy per unit mass and $a(1-V^2) |r|$ plays the role of the trapping gravitational potential.

To find a connection of the parameter $k$ with the proper time let us insert (\ref{yz}), (\ref{dx}), (\ref{dt}) and (\ref{dr}) into the definition of the interval (\ref{metric}),
\begin{equation}
ds^2 = e^{2a|r|} dt^2 - e^{2a|r|+u} dx^2  - dr^2 = \left[ (1+2a|r|) (1-V^2) - \epsilon \right]dk^2~.
\end{equation}
We see that on the brane ($r=0$) for photons $\epsilon = 1$ and for massive particles $0 < \epsilon < 1$.

From (\ref{dr}) it is clear that the motion towards the extra dimension $r$ is possible when
\begin{equation}
\epsilon - a (1-V^2) |r| \geq 0~,
\end{equation}
and for any energy $\epsilon$ there exists the maximal distance in the bulk,
\begin{equation}
|r|_{max} \sim \frac {\epsilon}{a}~,
\end{equation}
the particle can reach, i.e. the classical particles are trapped on the brane.

In the standard brane approach with decreasing warp factor ($a<0$) localization was achieved due to the fact that the extra space actually is finite \cite{brane-1,brane-2}. In our case the increasing of the brane warp factor, $e^{2a|r|}$, creates the potential well that confines particles. 

Another point is that we had neglected the influence of the oscillating exponents in  (\ref{metric}). In this approximation the anisotropy of the background metric for classical particles is hidden. However, anisotropic nature of the primordial metric (\ref{metric}) can be exhibited in cosmological solutions \cite{cosmology-1,cosmology-2}.

%%%%%%%%%%%%%%%%%%%%%%%%%%%%%%%%%%%%%%%%%%%%%%%%%%%%%%%%%%%%%%%

\section{The summary and discussions}

In this letter we have demonstrated the pure gravitational localization of all kinds of matter fields within the 5D standing wave braneworld \cite{Wave-1,Wave-2}. The main differences of our model from the standard brane approaches \cite{brane-1,brane-2} are:

i) Metric {\it ansatz} contains the increasing warp factor;

ii) In the bulk there exist rapidly oscillating standing waves which localize fields on the brane;

iii) The extra dimension factors of zero modes are not constant and far from the brane behave as:
\begin{eqnarray} \label{zeros}
\Phi (r) &\sim& e^{-4a|r|}~, \nonumber \\
g^{NM}A_M (r) &\sim&  e^{-4a|r|}~, \nonumber \\
\Psi_L (r) &\sim& e^{-3a|r|}~, ~~~~~ (r\to \infty) \\
\Psi_R (r) &\sim& e^{-2a|r|}~, \nonumber\\
h_{MN} (r) &\sim& e^{-4a|r|}~. \nonumber
\end{eqnarray}
where $\Phi$, $A^N$, $\Psi_L$, $\Psi_R$ and $h_{MN}$ correspond to the scalar, vector, left and right fermion and graviton respectively.

The advantage of our model is that the localization mechanism is universal, i.e. it should work also for interacting fields. If we suppose an extra dimensional profile for interacting fields as displayed in (\ref{zeros}), then it is clear that the integrals over $r$ of standard 5D interacting terms like $\sqrt{g} A^Nj_N$, $\sqrt{g} \Phi \overline{\Psi}_R\Psi_L$, $\sqrt{g} \Psi^2A^NA_N$, etc., are convergent and the extra dimensional factors only lead to renormalization of 4D coupling constants.

%%%%%%%%%%%%%%%%%%%%%%%%%%%%%%%%%%%%%%%%%%%%%%%%%%%%%%%%%%%%%%%%%%


\begin{thebibliography}{99}

\bibitem{Hi-1} N. Arkani-Hamed, S. Dimopoulos and G. Dvali,
            Phys. Lett. {\bf B 429} (1998) 263
            [hep-ph/9803315].

\bibitem{Hi-2} I. Antoniadis, N. Arkani-Hamed, S. Dimopoulos and G. Dvali,
            Phys. Lett. {\bf B 436} (1998) 257
            [hep-ph/9804398].

\bibitem{brane-1} M. Gogberashvili,
               Int. J. Mod. Phys. {\bf D 11} (2002) 1635
               [hep-ph/9812296];
               Mod. Phys. Lett. {\bf A 14} (1999) 2025
               [hep-ph/9904383].

\bibitem{brane-2} L. Randall and R. Sundrum,
               Phys. Rev. Lett. {\bf 83} (1999) 3370
               [hep-ph/9905221];
               Phys. Rev. Lett. {\bf 83} (1999) 4690
               [hep-th/9906064].

\bibitem{reviews-1} V.A. Rubakov,
                Phys. Usp. {\bf 44} (2001) 871 (Usp. Fiz. Nauk {\bf 171} (2001) 913).

\bibitem{reviews-2} D. Langlois,
                Prog. Theor. Phys. Suppl. {\bf 148} (2003) 181
                [hep-th/0209261].

\bibitem{reviews-3} P.D. Mannheim,
                {\it Brane-localized Gravity} (World Scientific, Singapore 2005).

\bibitem{reviews-4} R. Maartens and K. Koyama,
                Living Rev. Rel. {\bf 13} (2010) 5
                [1004.3962 [hep-th]].

\bibitem{S-1} M. Gutperle and A. Strominger,
           JHEP {\bf 0204} (2002) 018
           [hep-th/0202210].

\bibitem{S-2} M. Kruczenski, R.C. Myers and A.W. Peet,
           JHEP {\bf 0205} (2002) 039
           [hep-th/0204144].

\bibitem{S-3} V.D. Ivashchuk and D. Singleton,
           JHEP {\bf 0410} (2004) 061
           [hep-th/0407224].

\bibitem{S-4} C.P. Burgess, F. Quevedo, R. Rabadan, G. Tasinato and I. Zavala,
           JCAP {\bf 0402} (2004) 008
           [hep-th/0310122].

\bibitem{Wave-1} M. Gogberashvili and D. Singleton,
              Mod. Phys. Lett. {\bf A 25} (2010) 2131
              [0904.2828 [hep-th]].

\bibitem{Wave-2} M. Gogberashvili, A. Herrera-Aguilar and D. Malag\'on-Morej\'on,
              Class. Quantum Grav. {\bf 29} (2012) 025007
              [1012.4534 [hep-th]].

\bibitem{BaGa} B. Bajc and G. Gabadadze,
              Phys. Lett. {\bf B 474} (2000) 282
              [hep-th/9912232].

\bibitem{Po} A. Pomarol,
            Phys. Lett. {\bf B 486} (2000) 153
            [hep-ph/9911294].

\bibitem{Od} I. Oda,
            Phys. Rev. {\bf D 62} (2000) 126009
            [hep-th/0008012].

\bibitem{6D-1} M. Gogberashvili and P. Midodashvili,
            Phys. Lett. {\bf B 515} (2001) 447
            [hep-ph/0005298];
            Europhys. Lett. {\bf 61} (2003) 308
            [hep-th/0111132].

\bibitem{6D-2} M. Gogberashvili and D. Singleton,
            Phys. Lett. {\bf B 582} (2004) 95
            [hep-th/0310048];
            Phys. Rev. {\bf D 69} (2004) 026004
            [hep-th/0305241].

\bibitem{6D-3} M. Gogberashvili, P. Midodashvili and D. Singleton,
            JHEP {\bf 0708} (2007) 033
            [0706.0676 [hep-th]].

\bibitem{phantom-1} K. Bronnikov,
                 Acta. Phys. Pol. {\bf B 4} (1973) 251.

\bibitem{phantom-2} R.R. Caldwell,
                 Phys. Lett. {\bf B 545} (2002) 23
                 [astro-ph/9908168].

\bibitem{phantom-3} M. Pospelov,
                 Int. J. Mod. Phys. {\bf A 23} (2008) 881
                 [hep-ph/0412280].

\bibitem{phantom-4} A. Das, S. Kar and S. SenGupta,
                 Int. J. Mod. Phys. {\bf A 24} (2009) 4457
                 [0804.1757 [hep-th]].

\bibitem{phantom-5} V. Dzhunushaliev, V. Folomeev and M. Minamitsuji,
                 Rept. Prog. Phys. {\bf 73} (2010) 066901
                 [0904.1775 [gr-qc]].

\bibitem{GMS-1} M. Gogberashvili and R. Khomeriki,
             Mod. Phys. Lett. {\bf A 24} (2009) 2761
             [0808.1295 [gr-qc]].

\bibitem{GMS-2} M. Gogberashvili, S. Myrzakul and D. Singleton,
             Phys. Rev. {\bf D 80} (2009) 024040
             [0904.1851 [gr-qc]].

\bibitem{GMM-1} M. Gogberashvili, P. Midodashvili and L. Midodashvili,
               Phys. Lett. {\bf B 702} (2011) 276
               [1105.1701 [hep-th]].

\bibitem{GMM-2} M. Gogberashvili, P. Midodashvili and L. Midodashvili,
               Phys. Lett. {\bf B 702} (2011) 276
               [1105.1701 [hep-th]].

\bibitem{GMM-3} M. Gogberashvili, P. Midodashvili and L. Midodashvili,
               arXiv: 1109.3758 [hep-th].

\bibitem{cosmology-1} M. Gogberashvili, A. Herrera-Aguilar, D. Malag\'on-Morej\'on and R. Mora-Luna,
                   arXiv: 1202.1608 [hep-th].

\bibitem{cosmology-2} M. Gogberashvili, A. Herrera-Aguilar, D. Malag\'on-Morej\'on, R. Mora-Luna and U. Nucamendi,
                   arXiv: 1201.4569 [hep-th].


\end{thebibliography}
\end{document}